\documentclass[aps,twocolumn,showpacs]{revtex4}
\usepackage{graphicx}
\usepackage{amssymb}

\begin{document}

\title{Influence of stability islands in the recurrence of particles in a 
static oval billiard with holes}

\author{Matheus Hansen$^1$, R. Egydio de Carvalho$^2$ and Edson D.\ 
Leonel$^{3,4}$}
\affiliation{$^1$ Instituto de F\'isica da Universidade de S\~ao Paulo, Rua do 
Mat\~ao, Travessa R 187, Cidade Universit\'aria, 05314-970 S\~ao Paulo, SP - 
Brazil\\
$^2$ Universidade Estadual Paulista - UNESP, Rio Claro-SP, Brazil\\
$^3$ Departamento de F\'{\i}sica, UNESP - Univ Estadual Paulista, Av. 24A, 
1515, Bela Vista, 13506-900, Rio Claro, SP - Brazil\\
$^4$Abdus Salam International Center for Theoretical Physics, Strada  Costiera 
11, 34151 Trieste, Italy}

\date{\today} \widetext

\pacs{05.45.-a, 05.45.Pq, 05.45.Tp}

\begin{abstract}
Statistical properties for the recurrence of particles in an oval billiard 
with a hole in the boundary are discussed. The hole is allowed to move in the 
boundary under two different types of motion: (i) counterclockwise periodic circulation
with a fixed step length and; (ii) random movement around the 
boundary. After injecting an ensemble of particles through the hole we show that the 
surviving probability of the particles without recurring - 
without escaping - from the billiard is described by an exponential law and that 
the slope of the decay is proportional to the relative size of the hole. Since 
the phase space of the system exhibits islands of stability we show that there are 
preferred regions of escaping in the polar angle, hence given a partial answer to 
an open problem: {\it Where to place a hole in order to maximize or minimize a 
suitable defined measure of escaping}.
\end{abstract}
\maketitle

\section{Introduction}
\label{sec1}

A billiard is a dynamical system where a point-like particle moves with 
constant speed along straight lines confined to a piecewise and smooth 
boundary to where it experiences specular reflections \cite{Ref1}. In such 
type of collisions the tangent component of the velocity of the particle, 
measured with respect to the border where collision happened, is unchanged while 
the normal component reverses sign. Originally, the investigation on billiards 
was introduced in the seminal paper of Birkhoff \cite{Ref2} in the beginning 
of last century -- therefore introducing a new research area -- and since from 
there the scientific research on this topic has experienced a great development. 
Indeed, Birkhoff considered the investigation of the motion of a free point-like 
particle in a bounded manifold. Modern investigations on billiards 
however are connected with the results of Sinai \cite{Ref3} and Bunimovich 
\cite{Ref4,Ref5} who made rigorous demonstrations in the subject. The billiards 
theory has also been used in many different kinds of physical systems, 
including experiments on superconductivity \cite{Ref6}, wave guides 
\cite{Ref7}, microwave billiards \cite{Ref8,Ref9}, confinement of electrons in 
semiconductors by electric potentials \cite{Ref10,Ref12}, quantum tunneling \cite{Ref11} and 
many others.

The dynamics of a particle in a billiard can be matched into one of the 
following three 
possibilities: (i) regular; (ii) ergodic and; (iii) mix. The 
circular billiard is a typical example of case (i) since it is integrable 
thanks to the conservation of energy and angular momentum \cite{Ref1}. The 
elliptic billiard is also integrable and observables preserved are the energy 
and angular momenta about the two foci \cite{Ref13}. Case (ii) corresponds to systems containing 
zero measure stable periodic orbits, hence dominated by chaotic dynamics as the Bunimovich 
stadium \cite{Ref4,Ref14} as well as the Sinai billiard \cite{Ref3}. Finally the 
case (iii) includes billiards where the phase space has mixed dynamics 
therefore containing either regular dynamics characterized by periodic islands 
and/or invariant spanning curves and chaos as for instance, the annular
billiard \cite{Ref11}. In Ref. \cite{Ref13}, Sir Michael 
Berry discussed a family of billiards of the oval-like shapes. The radius in 
polar coordinates has a control parameter, ($\epsilon$), which leads to a 
smooth transition from a circumference with ($\epsilon=0$) -- hence integrable -- to 
a deformed form with $(\epsilon\ne0)$. For sufficiently small 
($\epsilon$), a special set of invariant spanning curves exists in the phase space 
corresponding to the so called whispering gallery orbits. They are orbits moving 
around the billiard, close to the border, with either positive (counterclockwise 
dynamics) or negative (clockwise dynamics) angular momentum. As soon as the 
parameter reaches a critical value \cite{Ref15}, the invariant spanning curves 
are destroyed as well as the whispering gallery orbits.

Billiards can also be considered in the context of recurrence of particles 
\cite{Ref16,Ref17}, particularly related to the Poincar\'e recurrence \cite{Ref16,Ref18}. 
The recurrence can be measured from the injection and hence from the escape of 
an ensemble of particles by a hole made in the boundary. The dynamics is made 
such that a particle injected through the hole is allowed to move inside the billiard 
suffering specular reflections with the boundary until it encounters the hole 
again. At this point the particle escapes from the billiard. The number of 
collisions that the particle has had till the escape is computed and another particle 
with different initial condition is introduced in the system. The dynamics is 
repeated until a large ensemble of particles is exhausted. The statistics of the 
recurrence time is then obtained. The known results are 
that for a totally chaotic dynamics, the survival probability -- probability that 
the particle survives without escaping through the hole -- is 
described by an exponential function \cite{Ref19}. When there are resonance islands, or 
Kolmogorov-Arnold-Moser (KAM) curves, to where the particles can be sticky around for long times, 
the dynamics is then changed to a slower decay conjectured in Ref. \cite{Ref18} to be 
described by a power law of universal scaling with an exponent from the order 
of $-2.57$. Therefore, 
as considered recently in a chapter book by Dettmann \cite{Ref19} who discusses some open 
problems in billiard with holes, a particular question was posed regarding to escape 
of particles: {\it Optimisation: Specify where to place a hole to maximize or 
minimize a suitable defined measure of escaping}.

In the current paper we discuss the recurrence of particles in an oval-like shaped 
billiard with a hole in the boundary and our main goal is to move a step further 
as an attempt to give a partial answer to the above question. The hole is 
allowed to move around the boundary under two different rules: (i) periodic and; 
(ii) randomly. In either cases, we define fixed places around the boundary to 
where the hole can be introduced. In the case (i) the hole moves 
counterclockwise under two circumstances. As soon as the particle is injected 
through the hole, its position moves if the particle escapes through it with 
less than $5$ collisions with the boundary. If the particles does not escape 
until $5$ collisions, it moves counterclockwise to a neighboring allowed 
position and wait until a escape or to more than other $5$ collisions. The billiard perimeter
is divided in 63 equally steps for the hole tour. This 
process repeats injecting and escaping particles until all the ensemble is 
exhausted. In the case (ii) the hole moves randomly around the boundary 
respecting the time of $5$ collisions. The survival probability, 
obtained from the recurrence time that the particle spent to escape, is accounted for 
a large ensemble of noninteracting particles. At each time of a escape, the 
polar angle and the angle of the trajectory particle are known, hence the 
corresponding position in the phase space where the escape happened is known as
well. Then a statistics of the density of particles that escaped from a given 
region of the 
phase space can be computed. We show that the density of escape measured in both 
polar angle as well as the angle of the particle's trajectory present peaks and 
valleys. The peaks are associated to the high density occupation in the phase 
space while the valleys are mostly linked to the periodic islands domain. Our 
results then give a partial answer to the above open question, at least for the 
oval billiard which has mixed phase space.

This paper is organized as follows. In Sec. \ref{sec2} we discuss the model 
and the equations that fully describe the dynamics of the system. The escape 
properties for the particles when the hole moves periodically around the 
boundary are made in Sec. \ref{sec3}. The survival probability for the 
particles when the hole moves randomly around the boundary is discussed in  
Sec. \ref{sec4} while our final remarks and conclusions are drawn in Sec.
\ref{sec5}.

\section{The static oval billiard}
\label{sec2}

We discuss in this section how to obtain the equations that fully describe 
the dynamics of the system. To start with, the radius of the boundary in polar 
coordinate is given by
\begin{equation}
R(\theta,\epsilon,p)=1+\epsilon\cos(p\theta),
\label{eq2.1}
\end{equation}
where $\theta$ is the polar coordinate, $\epsilon$ corresponds to a perturbation parameter 
of the circle and $p>0$ is an integer number. For $\epsilon=0$ the system is 
integrable. The phase space is foliated \cite{Ref1} and only periodic and 
quasi-periodic orbits are observed. For $\epsilon\ne0$ the phase space is mixed 
containing both periodic, quasi-periodic and chaotic dynamics. When $\epsilon$ 
reaches the critical value \cite{Ref15} $\epsilon_{c}={1/(1+p^{2})}$ 
the invariant spanning curves, corresponding to the whispering gallery orbits 
are destroyed and only chaos and periodic islands are observed. This happens when the 
boundary is concave for $\epsilon<\epsilon_{c}$ and is not observed for 
$\epsilon>\epsilon_{c}$ when the boundary exhibits segments that are convex.

\begin{figure}[t]
\includegraphics[width=1.0\linewidth]{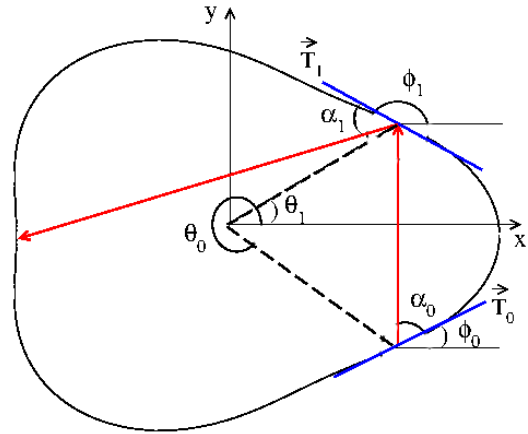}
\caption{(Color online). Illustration of the angles involved in the billiard.}
\label{Fig1}
\end{figure}

The dynamics is described by a two dimensional nonlinear mapping relating 
the variables $(\theta_n,\alpha_n)\rightarrow(\theta_{n+1},\alpha_{n+1})$ 
where $\theta$ denotes the polar angle to where the particle collides and 
$\alpha$ represents the angle that the trajectory of the particles does 
with a tangent line at the collision point. Fig. \ref{Fig1} illustrates the 
representation of the angles.

\begin{figure*}[t]
\begin{center}
\centerline{\includegraphics[width=8.5cm,height=5.5cm]{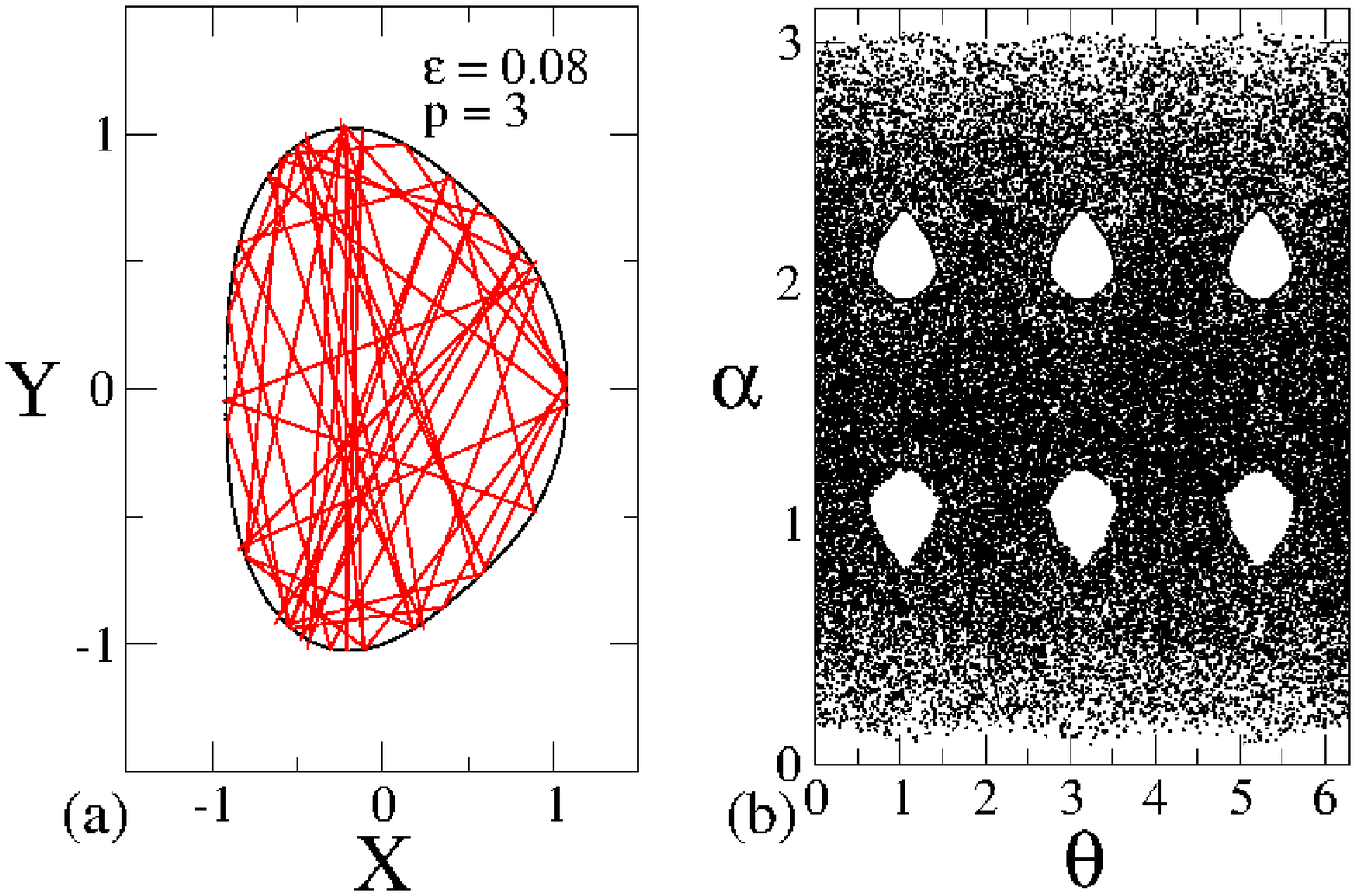}
            \includegraphics[width=8.5cm,height=5.5cm]{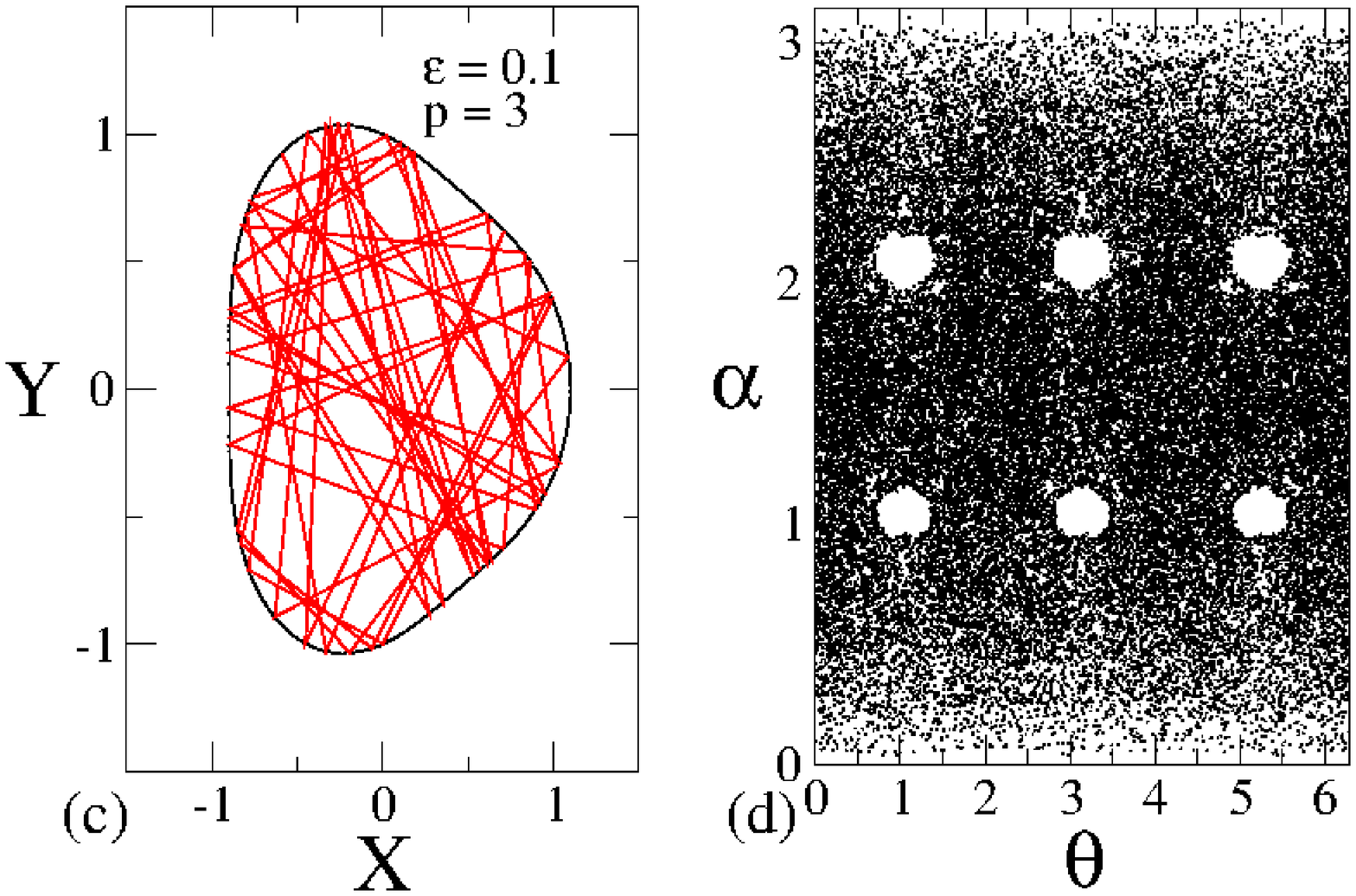}}
\vspace{0.5cm}
\centerline{\includegraphics[width=8.5cm,height=5.5cm]{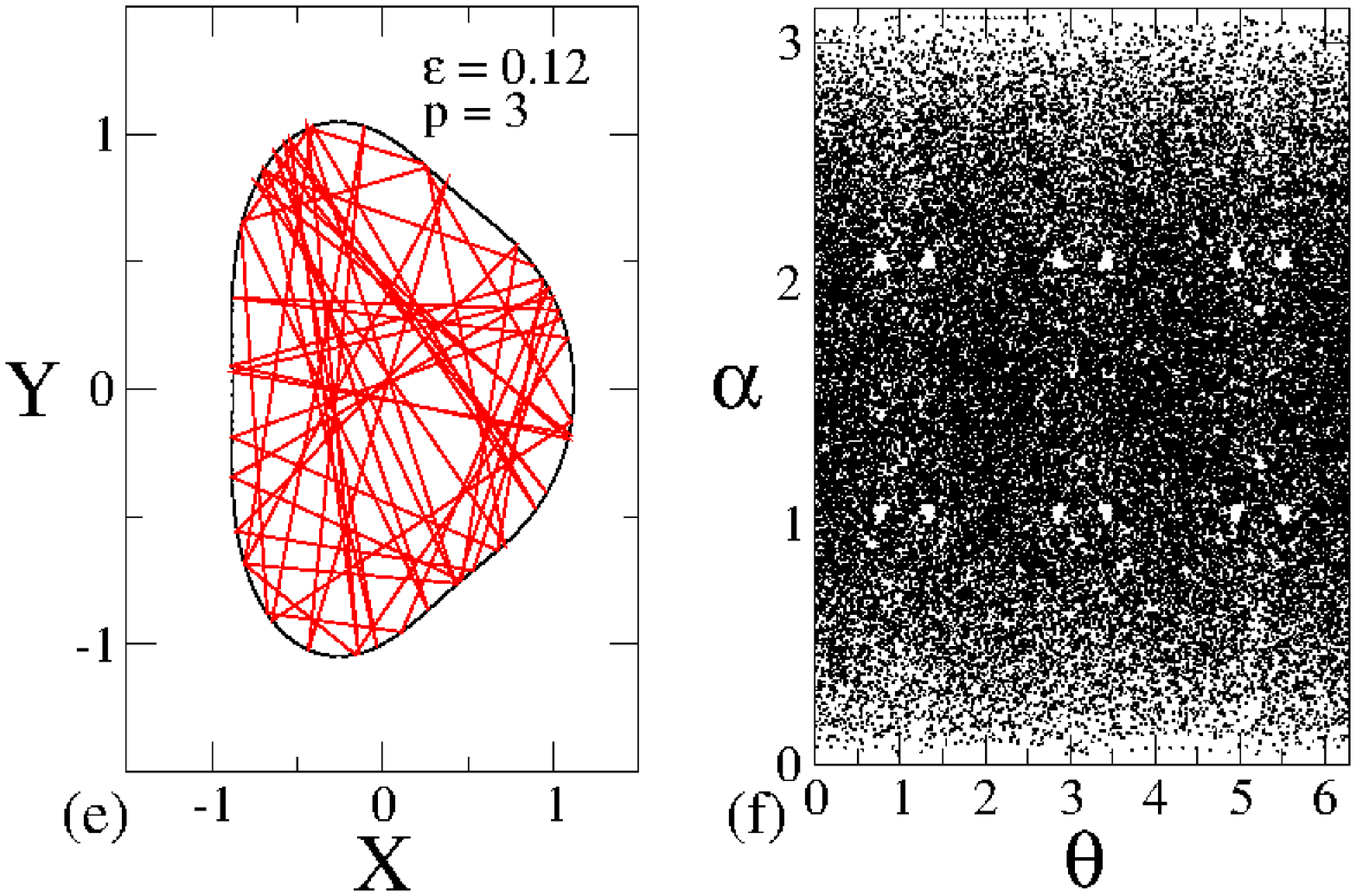}}
\end{center}
\caption{(Color online). Plot of the geometry shapes (a,c,e) and chaotic 
dynamics 
in such a boundaries in (b,d,f). The parameters used were $p=3$ and (a,b) 
$\epsilon=0.08<\epsilon_{c}$, (c,d) $\epsilon=0.1=\epsilon_{c}$ and (e,f) 
$\epsilon=0.12>\epsilon_{c}$.}
\label{Fig2}
\end{figure*}

For an initial condition $(\theta_n,\alpha_n)$ the position of the particle 
is written as $X(\theta_{n})=[1+\epsilon\cos(p\theta_{n})]\cos(\theta_{n})$ and 
$Y(\theta_{n})=[1+\epsilon\cos(p\theta_{n})]\sin(\theta_{n})$. The angle of the 
tangent vector at the polar coordinate $\theta_n$ is 
$\phi_{n}=\arctan\left[{Y^{\prime}(\theta_{n})\over{X^{\prime}(\theta_{n})}} 
\right]$, where $X^{\prime}(\theta)=dX(\theta)/d\theta$ and 
$Y^{\prime}(\theta)=dY(\theta)/d\theta$. Since there are no forces acting on the 
particle from collision to collision, it then moves along a straight line so 
its trajectory is given by
\begin{equation}
Y(\theta_{n+1})-Y(\theta_{n})=\tan(\alpha_{n}+\phi_{n})[X(\theta_{n+1}
)-X(\theta_{n})],
\label{eq2.5}
\end{equation}
where $\theta_{n+1}$ is the the new polar coordinate of the particle when it 
hits the boundary, which is to be obtained numerically. The angle 
$\alpha_{n+1}$ given the slope of the trajectory of the particle after a 
collision is
\begin{equation}
\alpha_{n+1}=\phi_{n+1}-(\alpha_{n}+\phi_{n}).
\label{eq2.10}
\end{equation}

The mapping is then written in a compact way as
\begin{equation}
\left\{\begin{array}{ll}
F(\theta_{n+1})&=R(\theta_{n+1})\sin(\theta_{n+1})-Y(\theta_{n})-\tan(\alpha_{n}
+\phi_{n})~\\ 
&\times[R(\theta_{n+1})\cos(\theta_{n+1})-X(\theta_{n})]~\\
\alpha_{n+1}&=\phi_{n+1}-(\alpha_{n}+\phi_{n})
\end{array} 
\right.~
\label{eq02.11}
\end{equation}
where $\theta_{n+1}$ is obtained numerically from $F(\theta_{n+1})=0$ with
$R(\theta_{n+1},\epsilon,p)=1+\epsilon\cos(p\theta_{n+1})$ and 
$\phi_{n+1}=\arctan 
\left[{Y^{\prime}(\theta_{n+1})}/{X^{\prime}(\theta_{n+1})}\right]$.

Figure \ref{Fig2}(a,c,e) represent the sketch of the boundary and few 
collisions of a typical chaotic orbit while Fig. \ref{Fig2}(b,d,f) correspond to the 
chaotic dynamics shown in the phase space. 
The parameters that we have used were $p=3$ and in Figures (a,b) $\epsilon=0.08<\epsilon_{c}$, in Figures (c,d) 
$\epsilon=0.1=\epsilon_{c}$ and in Figures (e,f) $\epsilon=0.12>\epsilon_{c}$.

\section{Statistical properties of the open oval billiard: the periodic case}
\label{sec3}

Let us now discuss some statistical properties for the recurrence of particles 
considering a hole of size $h$ placed on the boundary which can move regularly along 
the boundary with counterclockwise circulation. The hole has a constant aperture. Our 
initial results were obtained for a fixed $h=0.1$. Other 
values have been used too and the results are similar to the ones presented here. The 
hole is centered at $\theta_{ct}$ and we allow it to move around the 
boundary into $63$ possible places. The places are fixed and separated from each 
other by a step size of length $2\pi/63$. The dynamics of the particle is started 
from a hole, i.e., it is injected through it. The particle then moves around the 
boundary according to the equations of the mapping. If it happens that the 
particle visits the hole again within the interval of less than $5$ collisions, 
it escapes through the hole, a new and different initial condition is started 
and the process goes on. However, if the particle does not visit the same hole 
to where it was injected, the hole closes and opens a step above and has a time-life 
of $5$ collisions. If the particle does 
not reach such hole, it closes and reopens a step above with a life-time of $5$ 
collisions until the particle finds one hole to escape.

\begin{figure}[t]
\centering{\includegraphics[width=0.49\textwidth]{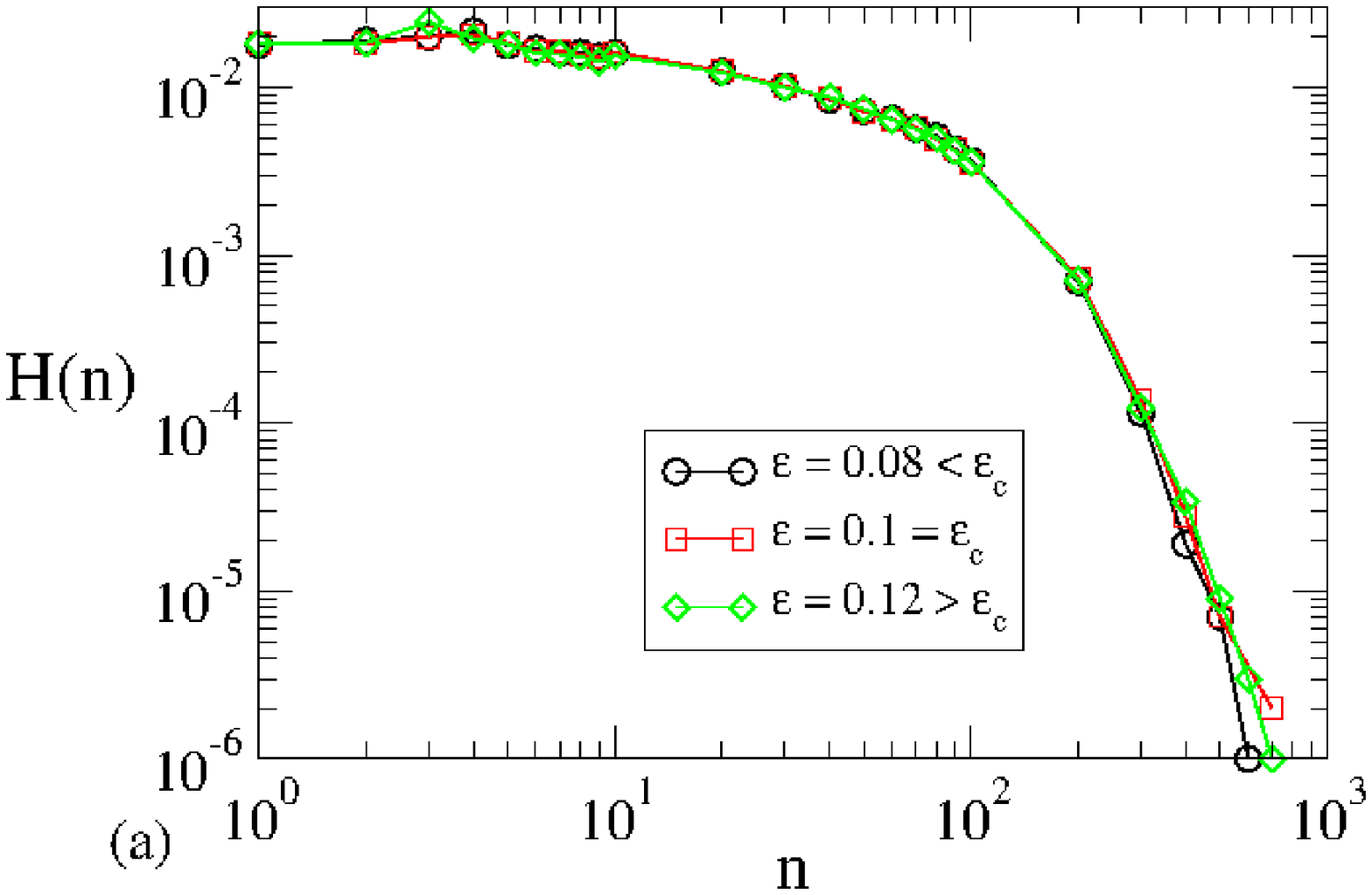}
           \includegraphics[width=0.49\textwidth]{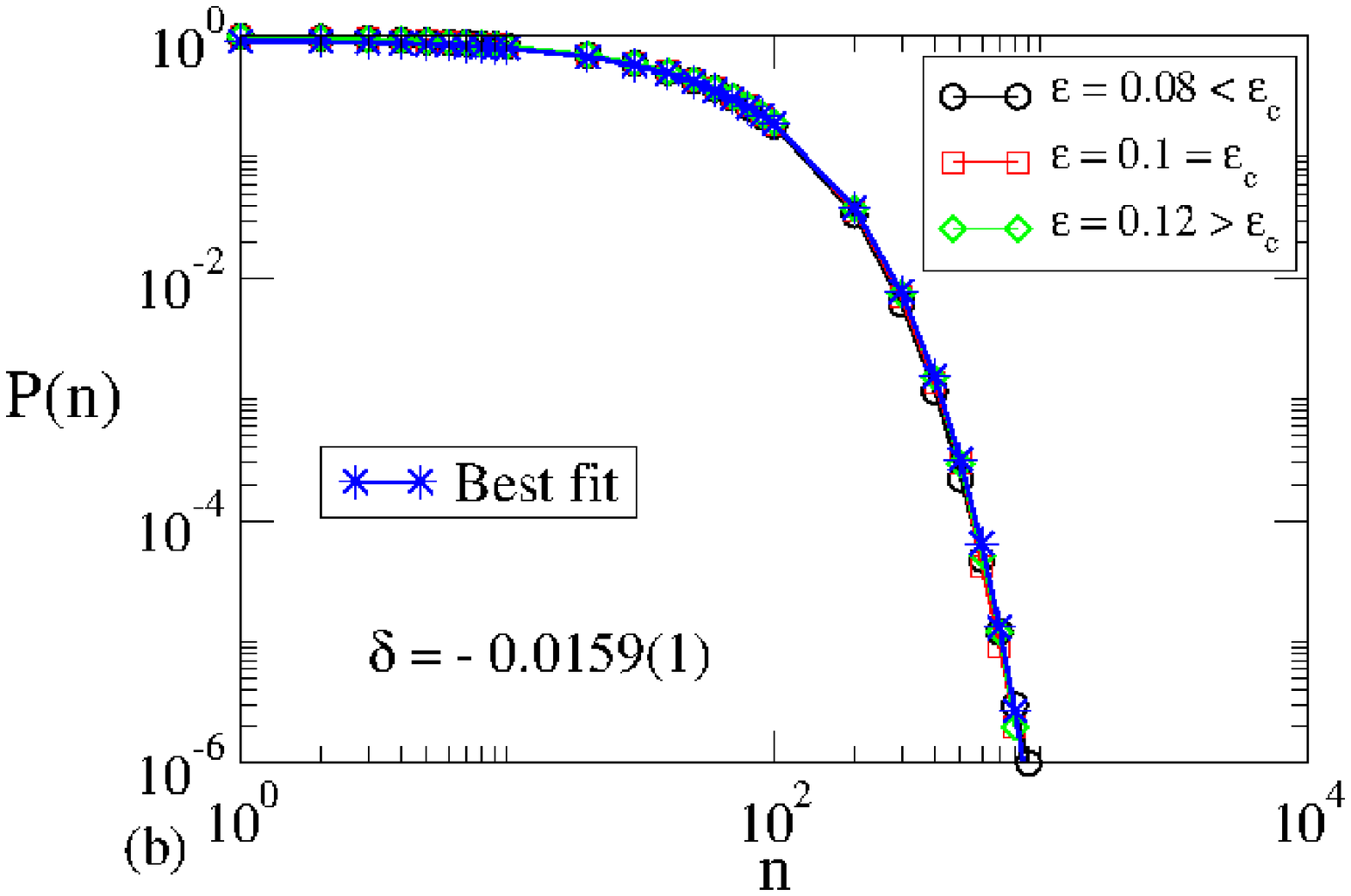}}
\caption{(Color online). (a) Histogram for the escape frequency of particles as 
a function of the number of collisions; (b) The survival probability of the 
particles. An exponential fitting furnishes $\delta=-0.0159(1)$. The parameter 
used was $p=3$ while the values of $\epsilon$ are shown in the plots.}
\label{Fig3}
\end{figure}

The procedure to study the escape of the particles and the survival probability 
is to inject an ensemble of $10^6$ particles from a hole placed in 
$\theta\in(0,h)$ and evolve the dynamics of each particle at most $10^6$ 
collisions with the boundary. The initial conditions were chosen such that 
$10^3$ of them were uniformly distributed in $\theta_{0}\in(0,h)$ and $10^{3}$ with different 
$\alpha_{0}\in(0,\pi)$. The statistical analysis were made in terms of 
the number of collisions of each particle until escape of the billiard. The process is 
repeated again until the whole ensemble is exhausted. Then a histogram showing the 
frequency of escapes is constructed. Fig. \ref{Fig3}(a) shows a histogram for the 
frequency of escape $H(n)$ for three different parameters, as labeled in the figure. The 
horizontal axis represents the number of collisions of the particle before reaching the 
hole and the vertical axis is the fraction of particles which escaped through a 
hole.

The integration of $H(n)$, gives the distribution of particles that do not 
escape through the hole until collision $n$. It corresponds to the survival 
probability and is given by 
\begin{equation}
P(n)={1\over{N}}\sum_{n=1}^{N}N_{surv}(n),
\label{eq2.12}
\end{equation}
where $N$ is the number of initial conditions and $N_{surv}$ is the number of the 
particles that survived until the $n^{th}$ collision. Fig. \ref{Fig3}(b) 
shows the results for the survival probability obtained for $p=3$ and for 
an ensemble of $10^6$ different particles. As one sees the decay law is 
exponential. We notice that after about $10^{3}$ collisions almost all initial 
conditions escaped from the billiard for a hole size $h=0.1$. The decay is 
fitted by a curve of the type
\begin{equation}
P(n)=P_{0}e^{n\delta}, 
\label{eq2.13}
\end{equation}
where $P_{0}$ is a constant, $\delta$ is the slope of the decay and $n$ is the 
number of collisions. The slope obtained by an exponential fit was 
$\delta=-0.0159(1)$, which is remarkably close to the relative size of the 
hole, i.e., the size of the hole over the whole length of the boundary
${h\over{2\pi}}={0,1\over{2\pi}}=0.0159\ldots$.

\begin{figure*}[t]
\centering{\includegraphics[width=0.32\textwidth]{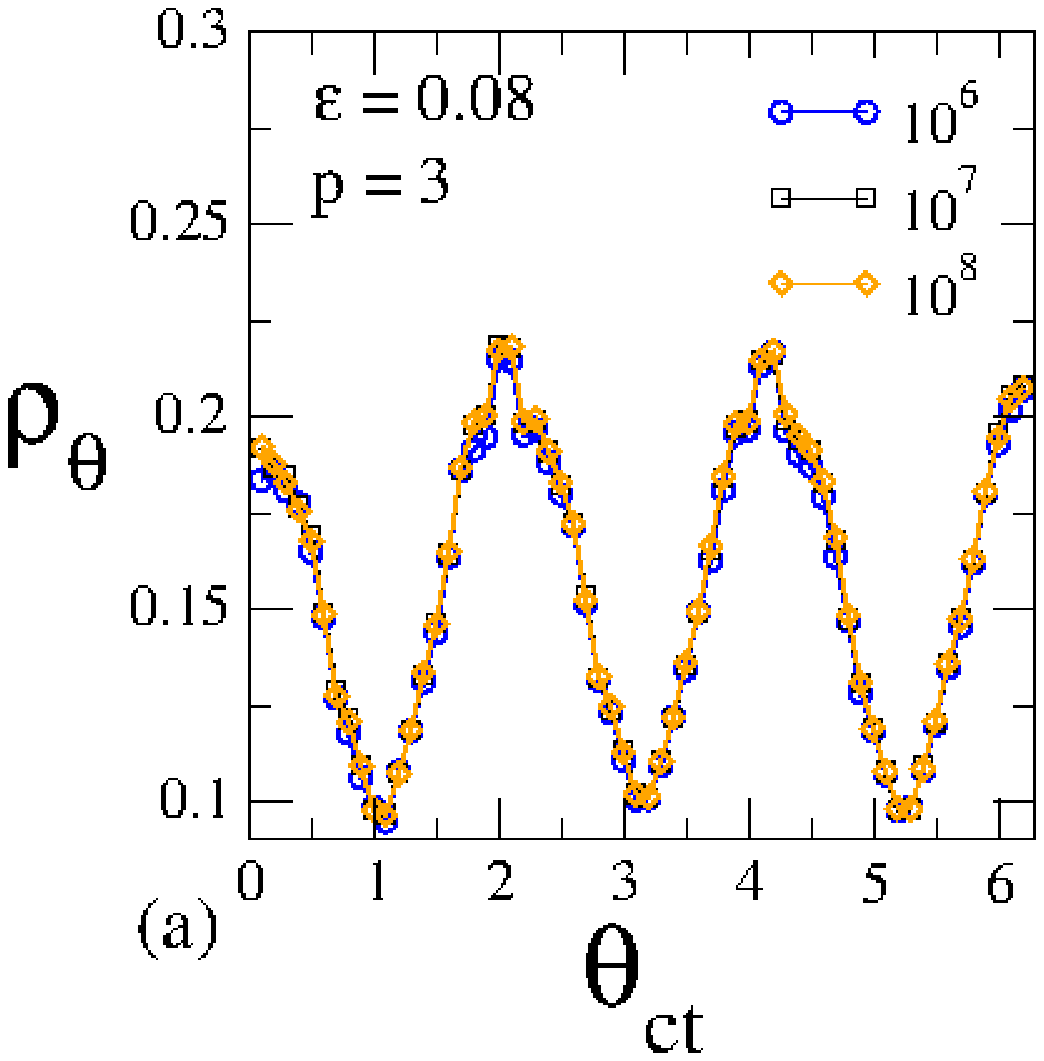}
           \includegraphics[width=0.32\textwidth]{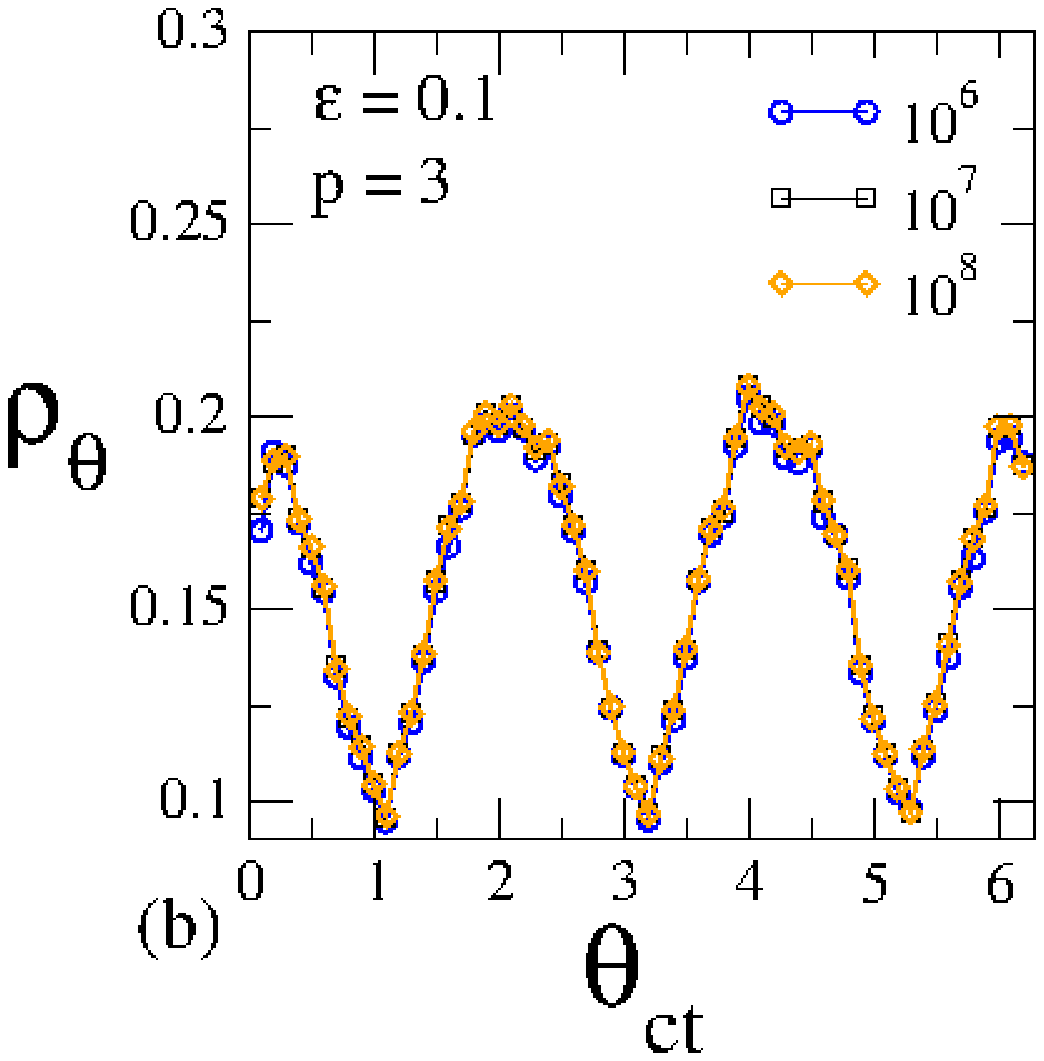}
           \includegraphics[width=0.32\textwidth]{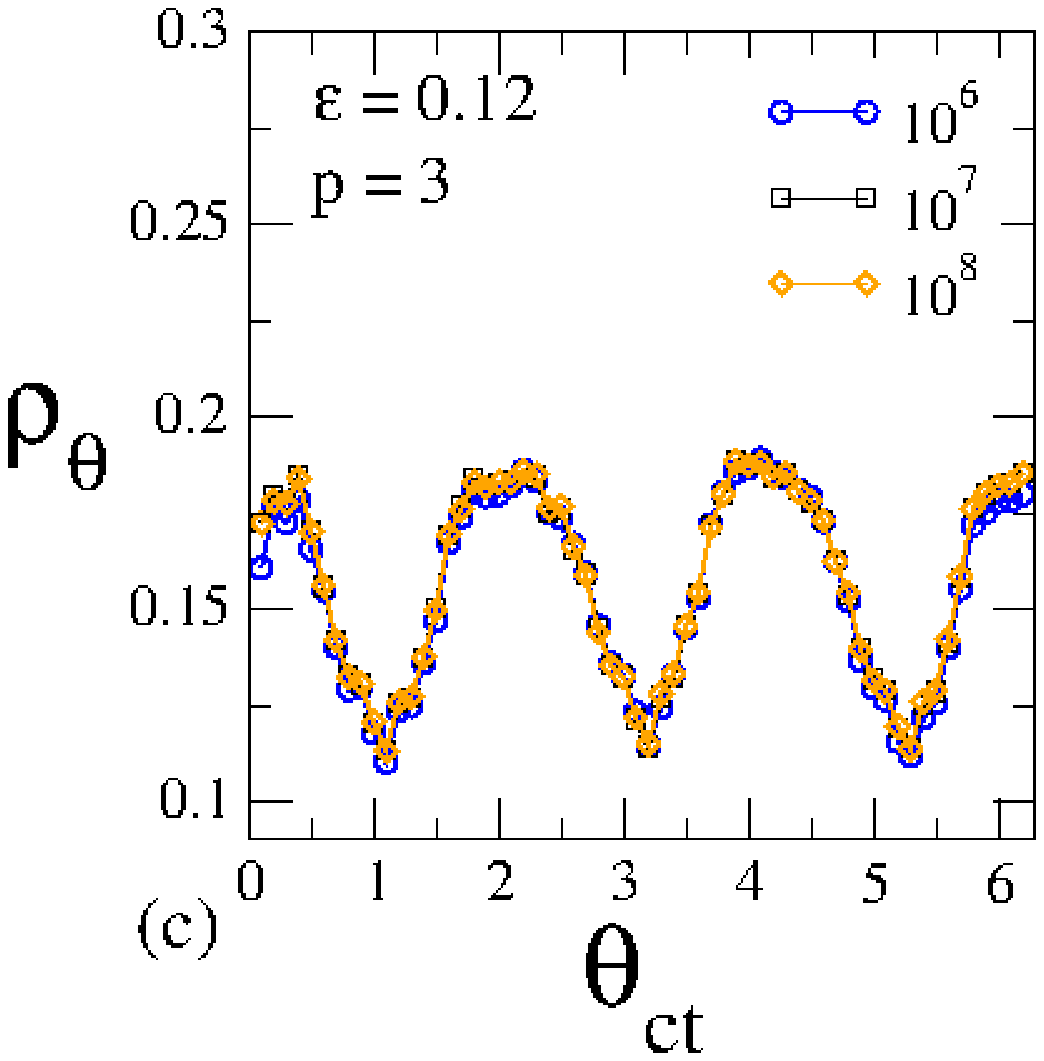}}
\centering{\includegraphics[width=0.32\textwidth]{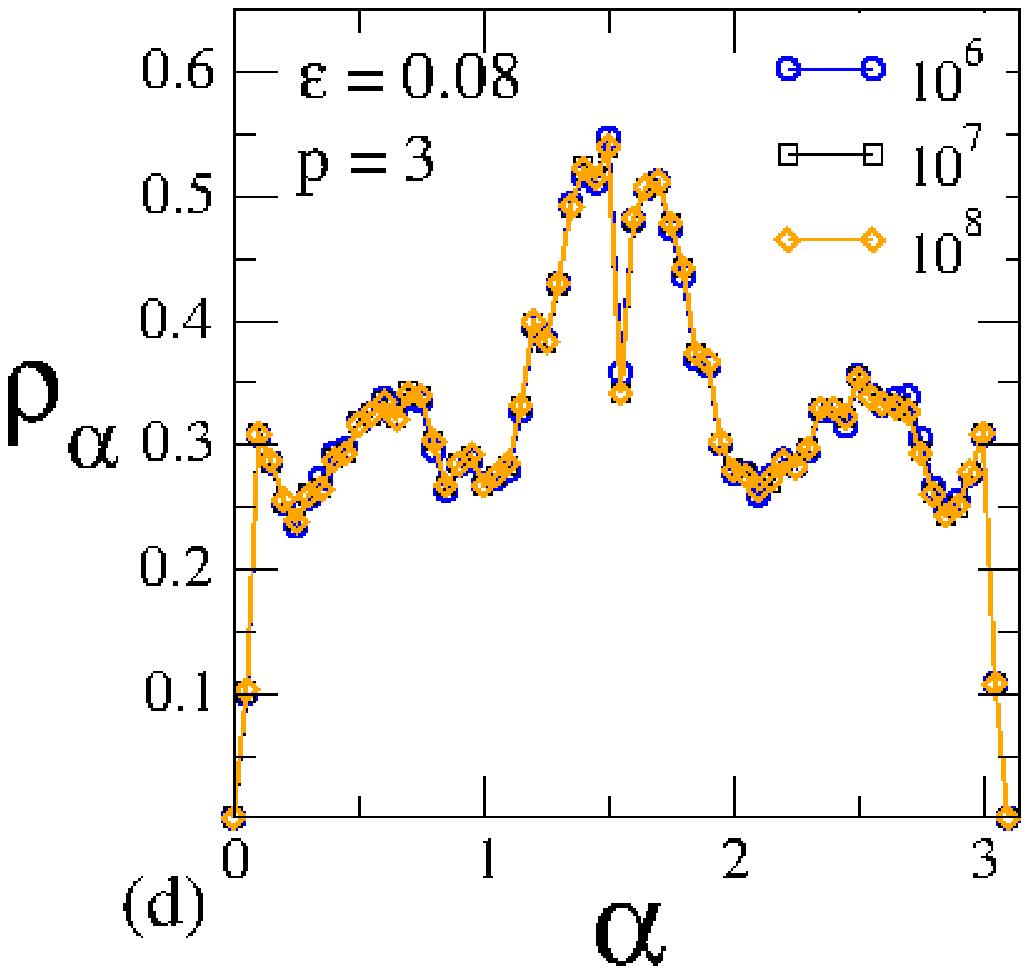}
           \includegraphics[width=0.32\textwidth]{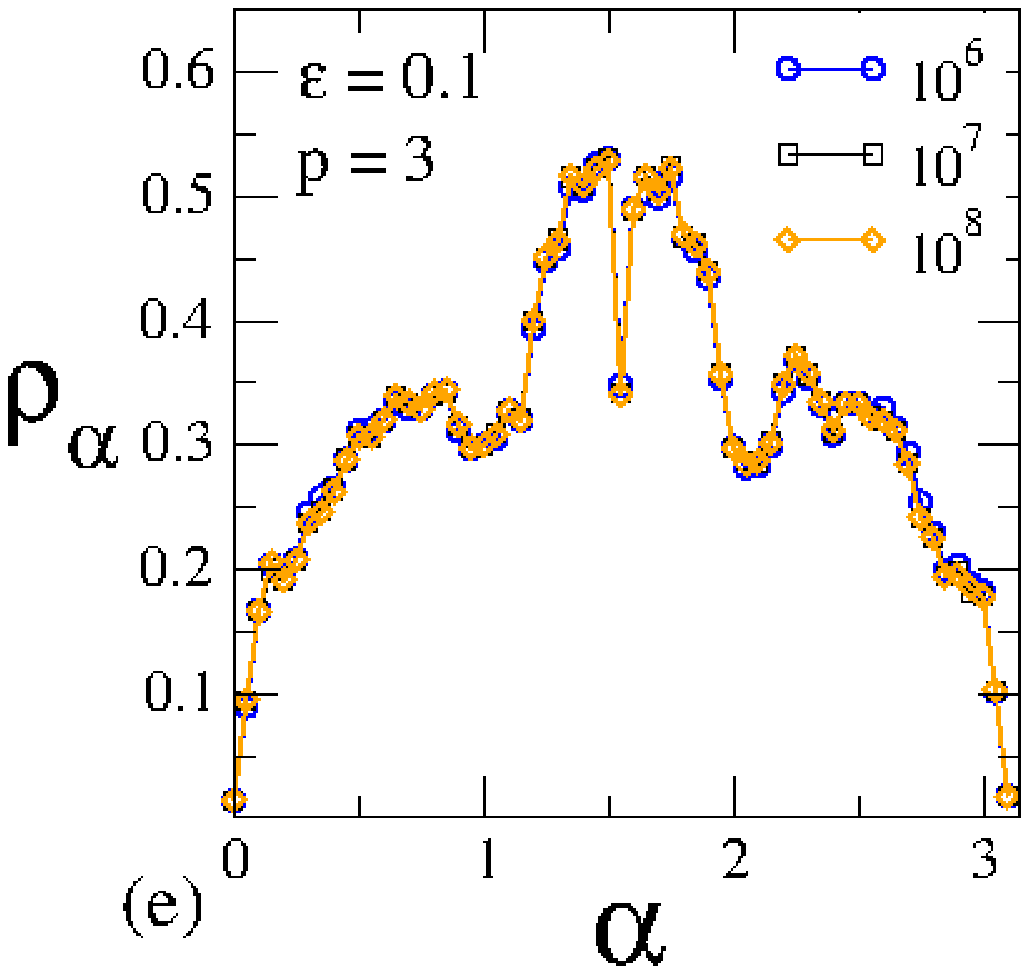}
           \includegraphics[width=0.32\textwidth]{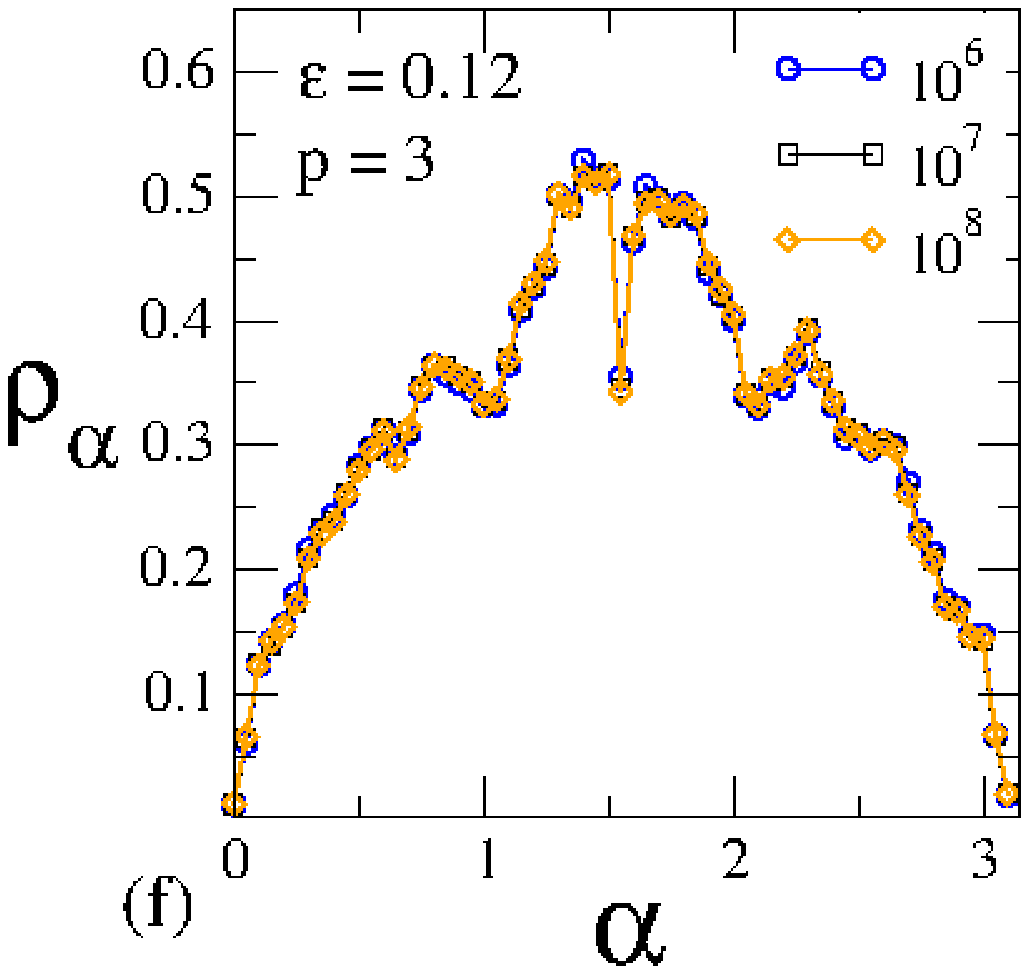}}
\caption{(Color online). Escaping density $\rho_{\theta}~vs.~\theta$ for the 
following parameters: (a) $\epsilon=0.08$; (b) $\epsilon=0.1$ and; (c) 
$\epsilon=0.12$. Three ensemble sizes were used, as labeled in the figures. The 
density of escaping $\rho_{\alpha}~vs.~\alpha$ is shown for: (d) $\epsilon=0.08$; 
(e) $\epsilon=0.1$ and; (f) $\epsilon=0.12$.}
\label{Fig4}
\end{figure*}

\begin{figure*}[!th]
\centering{\includegraphics[width=0.49\textwidth]{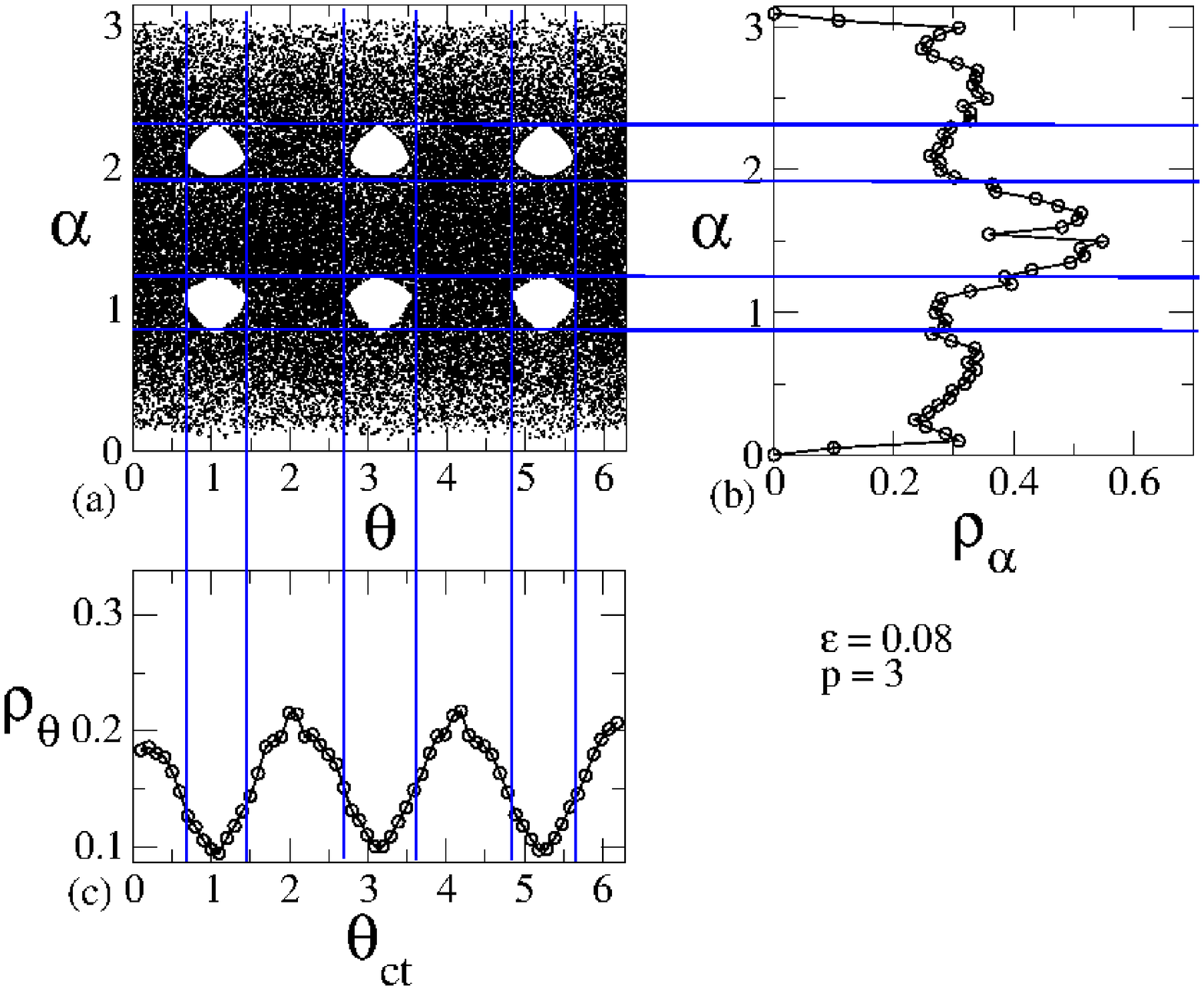}
           \includegraphics[width=0.49\textwidth]{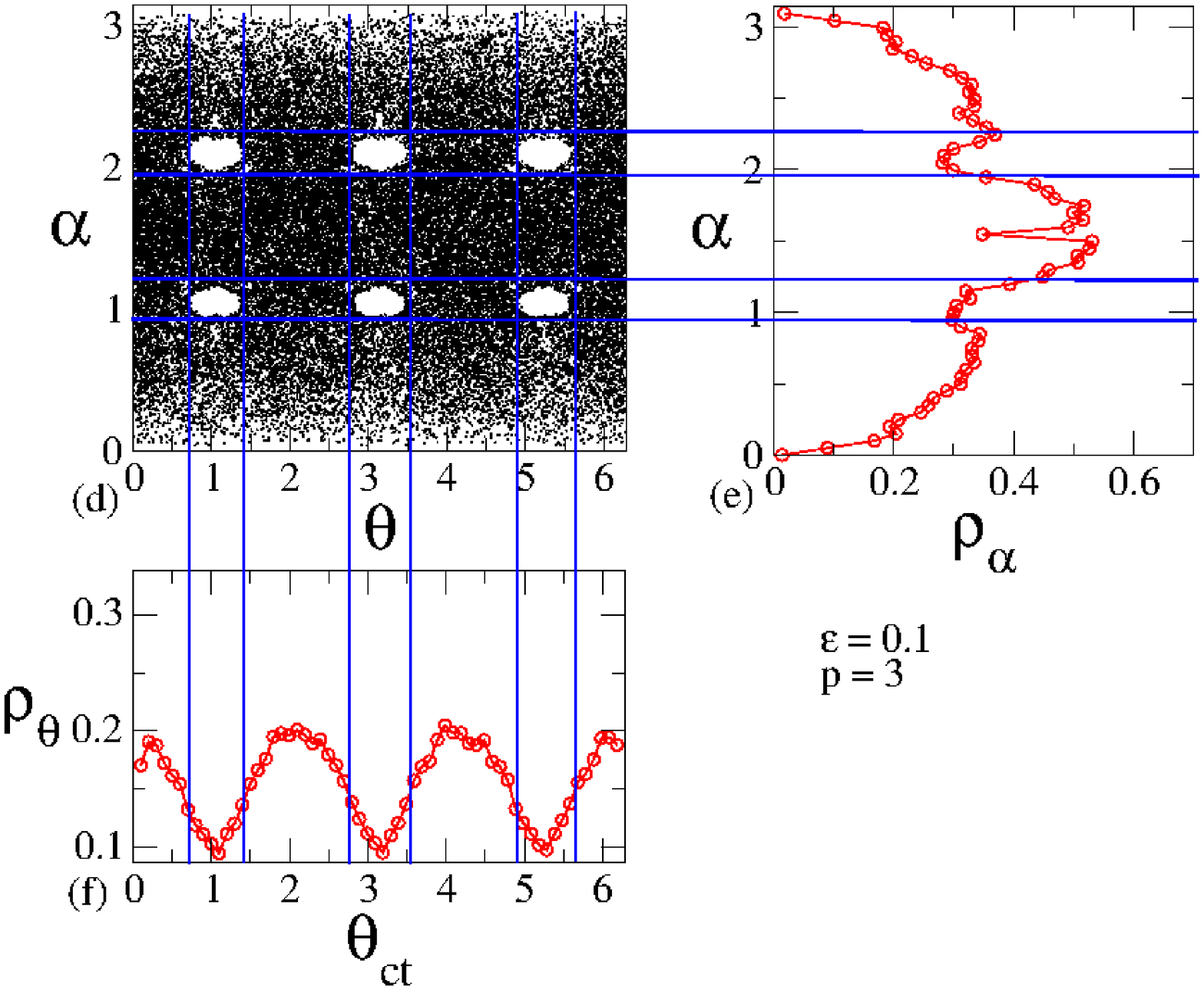}}
\centering\includegraphics[width=0.49\textwidth]{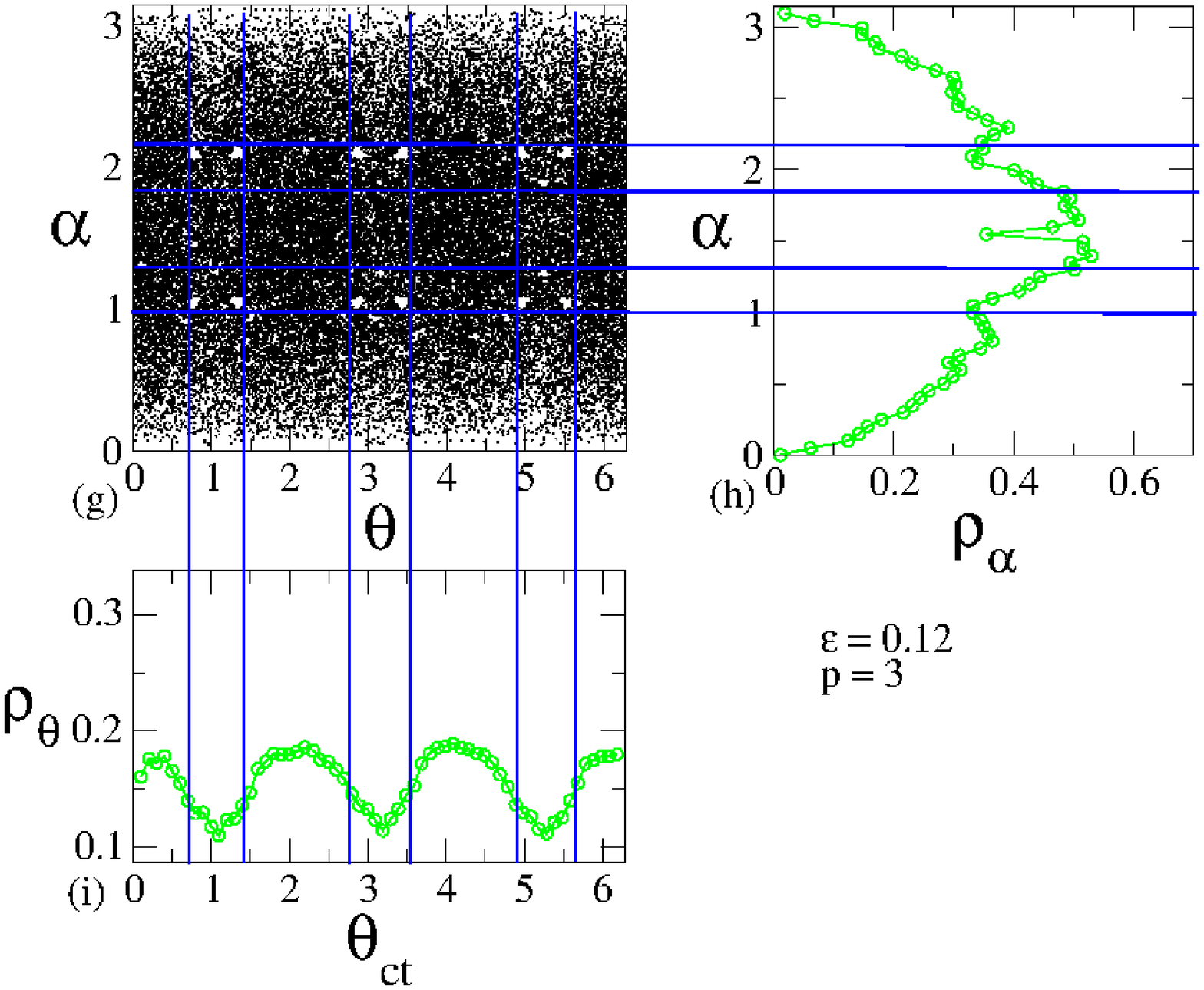}
\caption{(Color online). Connection of the peaks and valleys with the existence and 
absence of periodic regions in the phase space. The phase space is shown for: 
(a) $\epsilon=0.08$; (d) $\epsilon=0.1$ and; (g) $\epsilon=0.12$. 
The escape density $\rho_{\theta}~vs.~\theta$ is shown in; (b), (e) 
and (h) for the same combinations of control parameters.  The density 
$\rho_{\alpha}~vs.~\alpha$ is presented in: (c), (f), and (i).}
\label{Fig5}
\end{figure*}

Let us now discuss the density of escape measured as a function of the position 
of the hole $\theta_{ct}$. To do that we consider three different sizes for the 
ensemble of initial conditions, namely $10^{6}$, $10^{7}$ and $10^{8}$. Fig. 
\ref{Fig4} shows a plot of density $\rho_{\theta}~vs.~\theta_{ct}$ for the 
parameters: (a) $\epsilon=0.08$; (b) $\epsilon=0.1$ and; (c) $\epsilon=0.12$. We 
see from the figure the existence of peaks and valleys and that none difference 
of the curves for different ensemble sizes. The peaks correspond to the values of 
$\theta_{ct}$ to where the escape of particles is most probably to be observed, 
hence giving a clear evidence of preferable regions of escaping, while the valleys 
show the lower probability of escaping.

We may think of at first what is the reason that yields the density of escape 
to be smaller in some regions and larger in others. A possible answer for this 
could be linked directly to the periodic regions of the phase space. Then it 
becomes natural to look at the density of escaping also as a function of 
the variable $\alpha$. Fig. \ref{Fig4} (d-f) show the statistical results for the density 
of escape $\rho_{\alpha}~vs.~\alpha$ where $\alpha$ is the last coordinate the 
particle had prior the escape for the same parameters used in Fig. 
\ref{Fig4}(a-c). We see again that there are preferential regions to escape. 
Either peaks and valleys are also observed. The peaks and the valleys observed in both figures,
must be connected to each other and with the properties 
of the phase space. Thinking in this way Fig. \ref{Fig5}(a-i) show a connection of 
the valleys with the corresponding periodic regions in the phase space while the 
peaks are linked to the absence of periodic regions the phase space.

From the analysis of both Figs. \ref{Fig4} and \ref{Fig5} we can give a partial 
answer to the open question posed in \cite{Ref19}. The maximization of escape 
is provided at a region of the phase space with absence of periodic islands. In the complementary 
way, the minimization of escape is produced in a region of the phase space 
filled with islands of periodicity. This then reduces the effective size of the chaotic sea reducing the 
chances of a particle to escape if a hole is placed in the coordinates of periodic 
regions.

We must also mention a second and indirect phenomenon which is linked to the 
regions of periodic islands, particularly small islands and a possible channel 
among them, and that may lead to an extra life-time for the particles which is a 
phenomena called as stickiness \cite{Ref18}. Although a chaotic orbit can pass
very close to a periodic region, in whose domain there exist islands, the 
dynamics can be trapped locally avoiding visitation to other places in the phase 
space, hence giving extra life to particles trapped around such regions.

\section{Statistical properties of the open oval billiard: the random case}
\label{sec4}

In this section we discuss the results for the hole moving around the boundary 
but occupying random positions. The rule which controls the hole movement around 
the boundary is very similar to the previous section. However, instead of sequentially 
moving a step, the hole moves randomly along the $63$ possible places 
defined along the boundary. To chose a random position for placing the hole we used the so 
called RAN2 random number generator \cite{Ref22}.

The results obtained for this type of moving hole are remarkably
similar to those discussed in the previous section. For a hole size $h=0.1$, 
the slope of the decay of the survival probability obtained numerically is 
$\delta=-0.0163(1)$, very comparable to the results obtained in previous 
section.

\begin{figure}[t]
\centering{\includegraphics[width=0.475\textwidth]{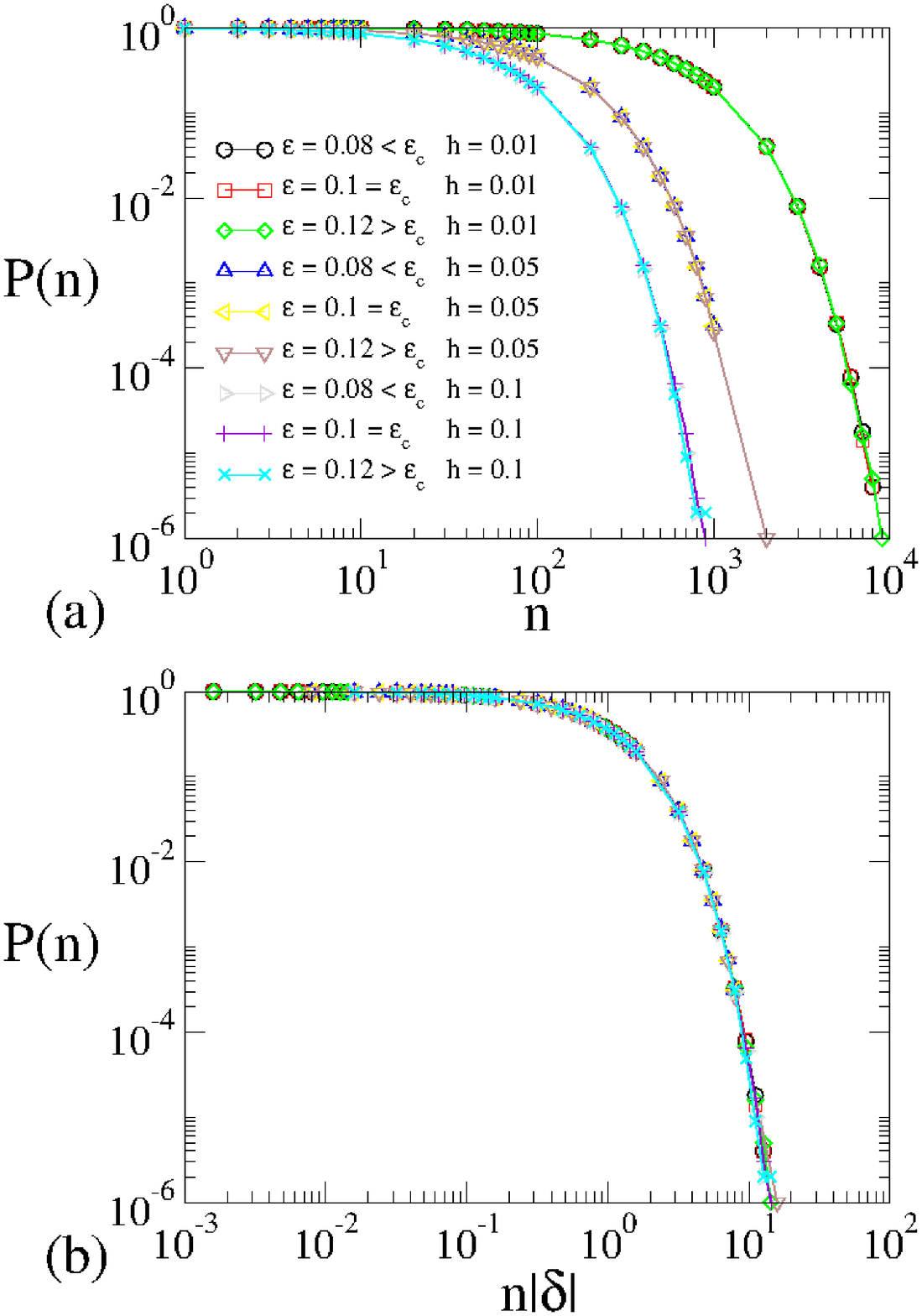}}
\caption{(Color online). (a) Plot of survival probability of the particles obtained for 
a hole moving randomly around the boundary as a function of $n$ for different 
hole sizes; (b) Their overlap onto a single and hence universal plot after a 
transformation $n\rightarrow n|\delta|$.}
\label{Fig6}
\end{figure}

Let us now discuss the influence of the hole size on the escape of 
particles. The exponential decay, marking the escaping dynamics of the 
particles does not change its pattern when the size of the hole is 
varied, at least for the range of holes considered here. For a smaller hole 
there is a shift in time allowing to the particles more opportunities to move along the 
phase space without escaping as compared to a larger hole. Therefore a scaling 
transformation $n\rightarrow n|\delta|$ rescales all curves obtained for the 
survival probability considering different values of $h$ onto a single and 
universal plot, as shown in Fig. \ref{Fig6}.

\section{Discussion and Conclusions}
\label{sec5}

In this paper we studied some statistical properties for the escaping and hence 
survival of particles inside an oval-like shaped billiard. We introduced a hole 
that changes its position on the boundary in two different ways: (i) first the 
hole moves around the boundary in a continuous stepwise either after a 
escape of a particle or after $5$ collisions with the boundary 
without finding the hole. In the second case the hole moves randomly after a 
escape or after a $5$ collisions life-time. For both cases the survival 
probability decays exponentially and the slope of the decay is proportional to 
the size of the hole over the total length of the boundary. 

We notice there are preferential regions along the phase space to where the 
escape of particles is facilitated, hence leading to a maximum escape. 
However, there are also regions in the space of phase, particularly those 
exhibiting stability islands, that influence quite negatively for the escape.

The discussion presented here allowed us to give a partial answer to the open 
question posed in \cite{Ref19}. We conclude that the maximization of escaping is 
obtained from a region of the phase space with absence of periodic islands 
and with higher density of visitation of particles. The opposite is also 
observed, the minimization of the escape can be produced from a region of the phase 
space filled with islands of periodicity. 
Stickiness can also provide with an extra life timefor the particles, by 
extending their dynamics over the chaotic sea.

\section{Acknowledgements}

MH thanks to CAPES for the financial support. REC and EDL acknowledge 
support from the Brazilian agencies FAPESP under the grants 2014/00334-9 and 
2012/23688-5 and CNPq through the grants 306034/2015-8 and 303707/2015-1.

\end{document}